\documentclass[twocolumn,superscriptaddress,preprintnumbers,aps, pre,amsmath,amssymb,floatfix,nofootinbib,notitlepage,showkeys]{revtex4-1}

\usepackage{graphicx}

\usepackage{color}

\usepackage{hyperref}

\begin{document}
\title{Active Flows and Deformable Surfaces in Development}

\author{Sami C.~Al-Izzi}
\affiliation{School of Physics and EMBL-Australia node in Single Molecule Science, University of New South Wales - Sydney 2052, Australia.}

\author{Richard G.~Morris}
\thanks{To whom correspondence should be addressed. E-mail: \texttt{r.g.morris@unsw.edu.au}.}
\affiliation{School of Physics and EMBL-Australia node in Single Molecule Science, University of New South Wales - Sydney 2052, Australia.}

\begin{abstract}
	We review progress in active hydrodynamic descriptions of flowing media on curved and deformable manifolds: the state-of-the-art in continuum descriptions of single-layers of epithelial and/or other tissues during development.  First, after a brief overview of activity, flows and hydrodynamic descriptions, we highlight the generic challenge of identifying the dependence on dynamical variables of so-called active kinetic coefficients--- active counterparts to dissipative Onsager coefficients.  We go on to describe some of the subtleties concerning how curvature and active flows interact, and the issues that arise when surfaces are deformable.  We finish with a broad discussion around the utility of such theories in developmental biology.  This includes limitations to analytical techniques, challenges associated with numerical integration, fitting-to-data and inference, and potential tools for the future, such as discrete differential geometry.
\end{abstract}
\maketitle
\tableofcontents

\section{Introduction}
\begin{figure*}
	\center\includegraphics[width=0.9\textwidth]{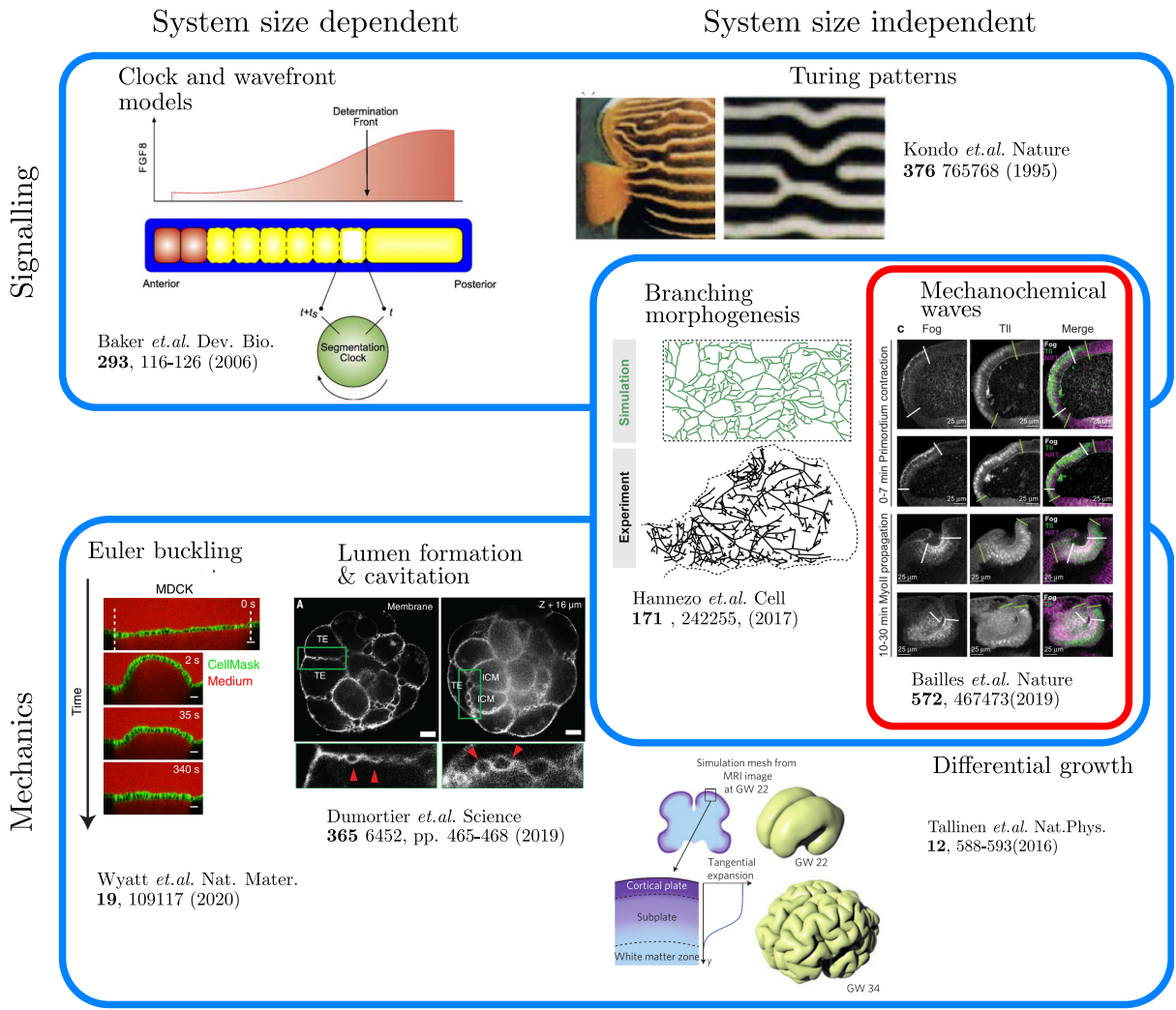}
		\caption{
			\textbf{A physicist's viewpoint on development}. A non-exhaustive schematic that arranges developmental studies according to whether their characteristic lengthscale(s) emerge due to predominantly mechanics or signalling (or both) and, moreover, whether those lengthscales are dependent or independent of the system size (\textit{i.e.}, if the system doubles in size, does the pattern double or the lengthscale itself, respectively).  Active flows on deformable surfaces are the state-of-the art for describing, analytically, the broad class of developing systems associated with single-cell thick epithelial sheets.  The characteristic lengthscales in such systems rely on both signalling and mechanics, and are typically size independent (although not always). The red box marks the type of processes that we will consider in this review. Figures reproduced with permission from \cite{baker_clock_2006}, \cite{kondo_reactiondiffusion_1995}, \cite{hannezo_unifying_2017}, \cite{bailles_genetic_2019}, \cite{wyatt_actomyosin_2020}, \cite{dumortier_hydraulic_2019}, \cite{tallinen_growth_2016}.
		}
	\label{fig:Fig1}
\end{figure*}

It is well documented that dramatic improvements to \textit{in vivo} imaging over the last $\sim$20 years has driven a resurgence in the field of developmental biology \cite{pantazis_advances_2014,mavrakis_lighting_2010}. However, this is not the only revolution to have taken place during such a time. Soft-condensed matter physics \cite{chaikin_principles_2000}, which emerged as a subject in its own right during the 1970s and 1980s, has also been irrevocably reshaped by relatively recent advances.

Ostensibly focussed on systems whose properties are characterised by energies of the order of $k_\mathrm{B}T$--- where $k_\mathrm{B}=1.38\times 10^{-23}$ $JK^{-1}$ is Boltzmann's constant and $T$ is temperature--- traditional soft matter \cite{doi_theory_1986,de_gennes_physics_1993,doi_soft_2013,safran_statistical_1994,seifert_configurations_1997} involves the study of passive materials, including gels, foams, emulsions, thin-films, membranes and ordered fluids. Due to their susceptibility on thermal energy scales, such materials are generically referred to as `soft', since they can be deformed relatively easily by applying macroscopic forces at their boundary--- \textit{e.g.}, imagine a piston compressing a fluid, or a spoon stirring an emulsion.

However, these ideas are somewhat at odds with a large and very prominent class of soft materials: that of \textit{living} systems. Here, energy (via metabolised nutrients) is provided \textit{locally}, taking the form of a constant (or externally controlled) chemical potential difference that drives microscopic forces on a molecular scale.  As a result, the deformability of living materials is more a question of whether such local energy input is of the same order, or greater, than the rate at which it can be dissipated, passively. 

We call such systems \textit{active} soft condensed matter.  Whilst active matter has developed very rapidly over the last 15-20 years \cite{marchetti_hydrodynamics_2013,ramaswamy_active_2017,julicher_hydrodynamic_2018} it still represents an area of intense research focus. Nevertheless, the judicious application of active theories has already raised the prospect that a broad range of biological systems--- previously considered beyond the remit of physics--- might be subject to theoretical treatments in a similar way that produced so much success in describing passive materials earlier in the last century. This is particularly true in the context of developmental biology, since active matter appears especially well-suited to the description of tissues and their behaviour during morphogenesis. 

Over one hundred years after d'Arcy Thompson's treatise \textit{``On Growth and From''} \cite{thompson_growth_1992}, physics is therefore back at the heart of development \cite{abzhanov_old_2017,lecuit_morphogenesis_2017,graner_forms_2017}. For the avoidance of doubt, in using the word \textit{physics}, the implication is not just that forces are at play. That much should be obvious, since controlling forces is the only way that a system can do work, and must therefore be central to all aspects of development. No, by using the word physics we are implicitly referring to one of the central tenets of the field: that of an \textit{effective} description, or model. The idea here is that, at the lengthscales of tissues and embryos, and on the timescales of observable deformations, the phenomenology of the system can be described by a model, or theory, that grossly oversimplifies the number of degrees-of-freedom at play.

Simple examples of successful effective theories from classical physics would be the Navier-Stokes equations that describe fluid dynamics, or the heat equation from non-equilibrium thermodynamics. Both have been empirically shown to accurately describe behaviours at long length- and time-scales, and neither require recourse to microscopic details.

In the context of developmental biology, most of the effective theories used so far can be classified according to a relatively simple criterion: the idea of lengthscale selection. That is, for a given aspect of morphogenesis, we ask: what are the dominant mechanisms by which characteristic lengthscales arise? This might be the wavelength of a buckled tissue, for instance, or the size of a cavity, or the mean separation between junctions in a branched system. In each of these examples, the underlying mechanisms are apparently quite different. However, at the highest level, we argue that only two principal axes suffice in order to organise different effective theories. The first is a putative spectrum that interpolates between signalling and mechanics, as the dominant method of lengthscale selection. The second is whether that lengthscale is dependent or independent, of the system's size. In Fig.~\ref{fig:Fig1} we have used this approach to organise a (non-exhaustive) selection of effective theories that are relevant in the context of developmental biology. These range from chemically driven lengthscale selection, in the form of clock and wavefront models \cite{cooke_clock_1976,baker_clock_2006} and Turing patterns \cite{turing_chemical_1952,kondo_reactiondiffusion_1995,maini_turings_2012}, to mechanically driven lengthscale selection from buckling \cite{wyatt_actomyosin_2020,nelson_buckling_2016,trushko_buckling_2020}, osmotic pressure \cite{dasgupta_physics_2018,dumortier_hydraulic_2019,verge_physics_2020} and differential growth \cite{goriely_differential_2005,goriely_mechanics_2015,tallinen_growth_2016,tozluoglu_planar_2019,hannezo_instabilities_2011}. Spanning these are a spectrum of mechanochemical processes, including: waves, invaginations and folds in epithelial sheets \cite{bailles_genetic_2019,mietke_self-organized_2019,miller_geometry_2018,collinet_programmed_2021}, and branching morphogenesis \cite{hannezo_statistical_2018,hannezo_unifying_2017}.

Amongst other things, such a classification helps to highlight what this short review is, and is not, about.  For example, there exist many good reviews concerning recent advances in developmental biology (\textit{e.g.}, \cite{lecuit_cell_2007,keller_physical_2012,heisenberg_forces_2013,gilmour_morphogen_2017,collinet_programmed_2021}) and there is no need to recapitulate them here.  Instead, we wish to focus on a particular sub-class of morphogenesis, that of \textit{active flows on deformable surfaces}. To understand why, it is first helpful to make a pedantic but important distinction: in the presence of active forces, flows can be generated in systems whose passive behaviour is either elastic or fluid, or a combination of both (or, indeed, something altogether more exotic).  That is, the mathematical concept of a flow \cite{frankel_geometry_2004}, and the physical concept of a fluid \cite{landau_fluid_2013} are not one in the same. In this context, active flows on deformable surfaces (or manifolds, as they are often referred-to) have a natural home in the description of a large class of processes in development; that of epithelial sheets and other single layer tissues.  For example, in the context of the model organism \textit{Drosophila melanogaster}, these include mechano-chemical waves observed in the midgut \cite{bailles_genetic_2019}, folds in fly wing \cite{tozluoglu_planar_2019}, and the formation of the hindgut \cite{nakamura_reduced_2013,nerurkar_molecular_2019}. In our classification scheme, such problems straddle both signalling and mechanics, with a lengthscale selection that is typically system-size independent (Fig.~\ref{fig:Fig1}).

We also limit our discussion to continuum treatments and so-called \textit{hydrodynamic} theories (Sec.~\ref{sec:hyd}), although we acknowledege a vast and sophisticated literature on \textit{in silico} representations of epithelial sheets and other tissues, often referred to as vertex models (\textit{e.g.}, \cite{rauzi_physical_2013,fletcher_vertex_2014,bi_motility-driven_2016,tetley_tissue_2019,rozman_collective_2020}). The reasons for this are threefold: firstly, vertex models are not our area of expertise; secondly, despite their drawbacks, we believe that analytical approaches provide a level of clarity that cannot be achieved by vertex models, and; thirdly, the role of geometry, and the subtleties involving its interplay with flows is poorly understood outside of a relatively small circle of aficionados.

The remainder of this article is structured as follows, we will first provide a very brief introduction to active hydrodynamics, in order to not only highlight some its the most salient features, but also to explain some of the major roadblocks to the construction of good effective theories.  We will go on to consider active hydrodynamic theories on deformable surfaces, describing the various aspects via which curvature and flows interact.  Finally, we will consider data-driven approaches, and the concomitant challenges of analysing data and constructing a theory, or model, that is consistent.

\section{Active Hydrodynamics}\label{sec:hyd}
So, what is a \textit{hydrodynamic} theory?  At the broadest level, a hydrodynamic theory captures the salient continuum behaviour of a system that emerges at macroscopic lengthscales and timescales.  Such theories manifest as (coupled) partial differential equations, where the variables are fields--- \textit{i.e.}, they depend on time and space.

Hydrodynamic descriptions are often called coarse-grained, since an ostensibly huge number of stochastic degrees-of-freedom have been averaged over, or ``integrated out''.  However, in most cases of practical interest, a formal coarse graining--- from a stochastic microscopic theory to a deterministic hydrodynamic one--- is fraught with technical challenges, rendering such an approach prohibitively difficult, if not impossible.  As a result, hydrodynamic theories are typically constructed phenomenologically, via recourse to symmetry principles or other heuristics gained from experimental observations.

One idea that is frequently leveraged is the notion that there are certain classes of fields whose behaviour in recpirocal space--- \textit{i.e.,} their Fourier modes--- is inherently long-wavelength and low-frequency \cite{hohenberg_theory_1977,chaikin_principles_2000}.  Of relevance to soft condensed matter are two such types of field:
\begin{enumerate}
	\item  \textit{Conserved} fields, including solute concentration, $c$, mass density, $\rho$, and momentum, $\rho\vec{v}$; and  
	\item  \textit{Broken symmetry} fields or Goldstone modes, such as a polarization field set up by chemical gradients, $\vec{P}$, or a nematic order parameter describing rod-like alignment, $\vec{n}=-\vec{n}$.
\end{enumerate}
For example, consider the momentum and mass conservation equations of an incompressible Navier-Stokes fluid \cite{landau_fluid_2013}:
\begin{align}
	&\rho\left(\partial_t\vec{v} + \vec{v}\cdot\nabla\vec{v}\right)=-\nabla p +\eta\nabla^2 \vec{v},\label{eq:NS}\\
	&\nabla\cdot\vec{v}=0.
	\label{eq:incomp}
\end{align}
By virtue of incompressibility, the pressure, $p$, can be eliminated. The remaining field, $\vec{v}$, is conserved since the density, $\rho$, is constant, implying that $\vec{v}$ is a proxy for momentum. More complex hydrodynamic theories are found throughout soft matter, in particular visco-elastic and polymeric gels \cite{doi_theory_1986}, various types of liquid crystal hydrodynamics \cite{de_gennes_physics_1993}, liquid crystal elastomers \cite{warner_liquid_2007}, colloidal suspensions with non-Newtonian behaviour \cite{poon_colloidal_2015,wyart_discontinuous_2014}, and the hydrodynamics of membranes and interfaces \cite{arroyo_relaxation_2009} to name but a few.

In passive hydrodynamic theories, the molecular details are essentially captured by kinetic coefficients, such as the viscosity, $\eta$, which appears in Eq.~(\ref{eq:NS}).  Such coefficients are typically constant, and do not vary in either space or time, although this is not always the case (\textit{e.g.}~shear thickening fluids \cite{gillissen_constitutive_2019}).  They have their values limited by the Onsager reciprocal relations \cite{onsager_reciprocal_1931-1,onsager_reciprocal_1931-2} and the Curie principle \cite{groot_non-equilibrium_1984}, which not only ensures that the fluctuation dissipation theorem (FDT) holds but also that, in the absence of external driving, the system will relax to thermodynamic equilibrium. 

By contrast, active systems, as set out briefly in the introduction, are characterised by a continual input of energy at a local scale, which is typically converted into forces in order to drive motion and/or deformation \cite{ramaswamy_mechanics_2010}. They are inherently out of equilibrium, and can reach steady states that break detailed balance and thus violate FDT.  In a hydrodynamic treatment, this manifests as active kinetic coefficients--- counterparts to the dissipative coefficients of passive theories--- that are \textit{not} constrained by the Onsager reciprocal relations nor the Curie principle \cite{groot_non-equilibrium_1984}. That is, active coefficients can give rise to negative entropy production, since they depend on both hidden degrees of freedom and also hidden energy consumption \cite{ramaswamy_active_2017}. Although ostensibly straightforward, it is these modifications that have led a revolution of sorts, giving rise to theories that have proved relevant across a broad range of lengthscales.  For example, applications have included, but are not limited to, the flocking of animals \cite{toner_flocks_1998}, chemically-driven self-motile particles \cite{fily_athermal_2012}, active liquid crystals \cite{ramaswamy_active_2003}, active membranes \cite{ramaswamy_nonequilibrium_2000}, and the cell cytoskeleton \cite{kruse_asters_2004,kruse_generic_2005,koster_cortical_2016,wollrab_polarity_2019,banerjee_actin_2020}.

In order to fix ideas, it is helpful to consider one of the canonical applications of active hydrodynamics: actin and myosin in the cortical cytoskeleton.  Here, myosin-II molecules oligomerise into mini-filaments that exert contractile forces on the underlying network of actin filaments.  In this case, a simple isotropic active contribution to the stress tensor takes the form \cite{marchetti_hydrodynamics_2013,julicher_hydrodynamic_2018}
\begin{equation}
	\label{eq:activeStressContractile}
	\sigma^{(a)}_{ij}= -\zeta\left(\rho_m,\Delta\mu_\mathrm{ATP},\ldots\right) g_{ij},
\end{equation}
where $g_{ij}$ is the metric tensor. The pre-factor $\zeta$ is an example of an active kinetic coefficient, controlling both the magnitude and sign of the resulting stresses: $\zeta<0$ gives a contractile stress whereas $\zeta>0$ gives an extensile stress.  Rather than being constant, as a passive kinetic coefficient would be, $\zeta$ can vary in both time and space due to its dependence on other degrees of freedom, such as myosin density, $\rho_m$, or the chemical potential of ATP hydrolysis, $\Delta\mu_\mathrm{ATP}$. Indeed, more biologically accurate descriptions must further recognise that myosin-II driven contractility involves myriad signalling pathways and feedback mechanisms.  For example, those surrounding the canonical upstream regulator RhoA \cite{nishikawa_controlling_2017,budnar_anillin_2019}, a small GTPase that regulates both myosin activation and the polymerisation of actin by formins.  In the context of development, where contractility shapes cells and hence tissues, relevant signalling pathways also include transcriptional regulation, which can produce mechano-chemical feedback on timescales of minutes and hours. For example, spatial profiles of expression are known to translate into patterns of signalling and hence characteristic tissue deformations \cite{bailles_genetic_2019}.

At this point, we remark that although Eq.~(\ref{eq:activeStressContractile}) suffices for the needs of pedagogy, additional anisotropic terms proportional to broken symmetry fields are also allowed ($\sigma^{(a)}_{ij}\sim P_{i}P_{j}$) \cite{marchetti_hydrodynamics_2013}, along with monopolar like force densities that are traditionally forbidden in passive hydrodynamics ($f^i\sim P^i$) \cite{maitra_activating_2014}.
Such terms are particularly pertinent to the dynamics of some \textit{in vitro} tissue culture assays \cite{saw_topological_2017,notbohm_cellular_2016,peyret_sustained_2019}, where nematic order can emerge from an initially isotropic monolayer \cite{mueller_emergence_2019}, and growing rod-like cell colonies \cite{doostmohammadi_defect-mediated_2016}. Often referred to as active nematic theories, approaches of this type are characterised by their ability to spontaneously generate topological defects which move dynamically through the fluid \cite{giomi_defect_2014}, forming turbulent-like states at sufficiently high activity \cite{thampi_active_2016}. In addition to this, extensile materials (a category which is believed to include at least some cell monolayers) undergo spontaneous shear-banding instabilities from initially ordered states \cite{duclos_spontaneous_2018}.  Nevertheless, the coefficients associated with all such active terms still suffer from the same issues as set out for the case of contractility. 

Most generally, we have a situation whereby \textit{physics} is captured by hydrodynamic equations, but \textit{biochemistry} dynamically controls the coefficients in those equations. This presents something of a problem: if active coefficients are, in fact, arbitrary functions of dynamical (and perhaps even hidden) degrees-of-freedom, then how can we know what form they take?  In principle, the answer is to follow standard scientific protocol. That is, make a hypothesis, compare the results with experimental observations, and repeat until good agreement is met.  However, in practice, biological materials are often very complex, with different types of active stresses intertwined with non-trivial passive responses.  Since the stresses that can be inferred by rheological and other experiments are typically combined, or net, quantities, this makes it hard to disambiguate between active contractility, for instance, and, say, passive elastic response.

This is not to say that all is lost.  On the contrary, such issues are surmountable with better experimental design and more extensive biological controls.  The challenge is therefore to highlight such issues so that experimental researchers are cognisant of potential pitfalls. Of particular note is recent progress in fitting active polar theories using the dynamics of constrained topological defects \cite{blanch-mercader_quantifying_2021,blanch-mercader_integer_2021}.

\section{Flows on Deformable Surfaces}

In the context of early-embryo development, the archetypal tissues responsible for morphogenesis are aggregates of epithelial cells, typically arranged in thin sheets \cite{guillot_mechanics_2013}. One of the most striking examples, and a system subject to intense experimental research, is the process of gastrulation in the model organism \textit{Drosophila melanogaster} \cite{sweeton_gastrulation_1991}.  Here  the  epithelium--- a  monolayer  of  cells,  tightly connected  to  each  other  via  proteins  such  as  E-cadherin---  undergoes embryo-scale deformations, including growth, invaginations and topological changes, which form the basis of the fly’s anatomy \cite{alberts_molecular_2002}.

Such complex shape behaviours require modifications to the equations of active matter, not only to incorporate the impact of curved surfaces on active flows, but also how active flows deform surfaces.  Here, we are reminded of a quote from the famous physicist, J.~A.~Wheeler, regarding Einstein's theory of general relativity:

\vspace{2mm}
\noindent\textit{``Space-time tells matter how to move; matter tells space-time how to curve.''}
\vspace{2mm}

\noindent With significant artistic interpretation, once can draw parallels with the equations that arise in morphogenesis.  That is, they describe degrees-of-freedom whose behaviour is coupled with that of the manifold on which those degrees-of-freedom are defined. In the following we touch on some of the complexities that arise in these cases, and the particular challenges involved in developing biologically relevant, yet tractable models.

\begin{figure*}[!htp]
\center\includegraphics[width=0.9\textwidth]{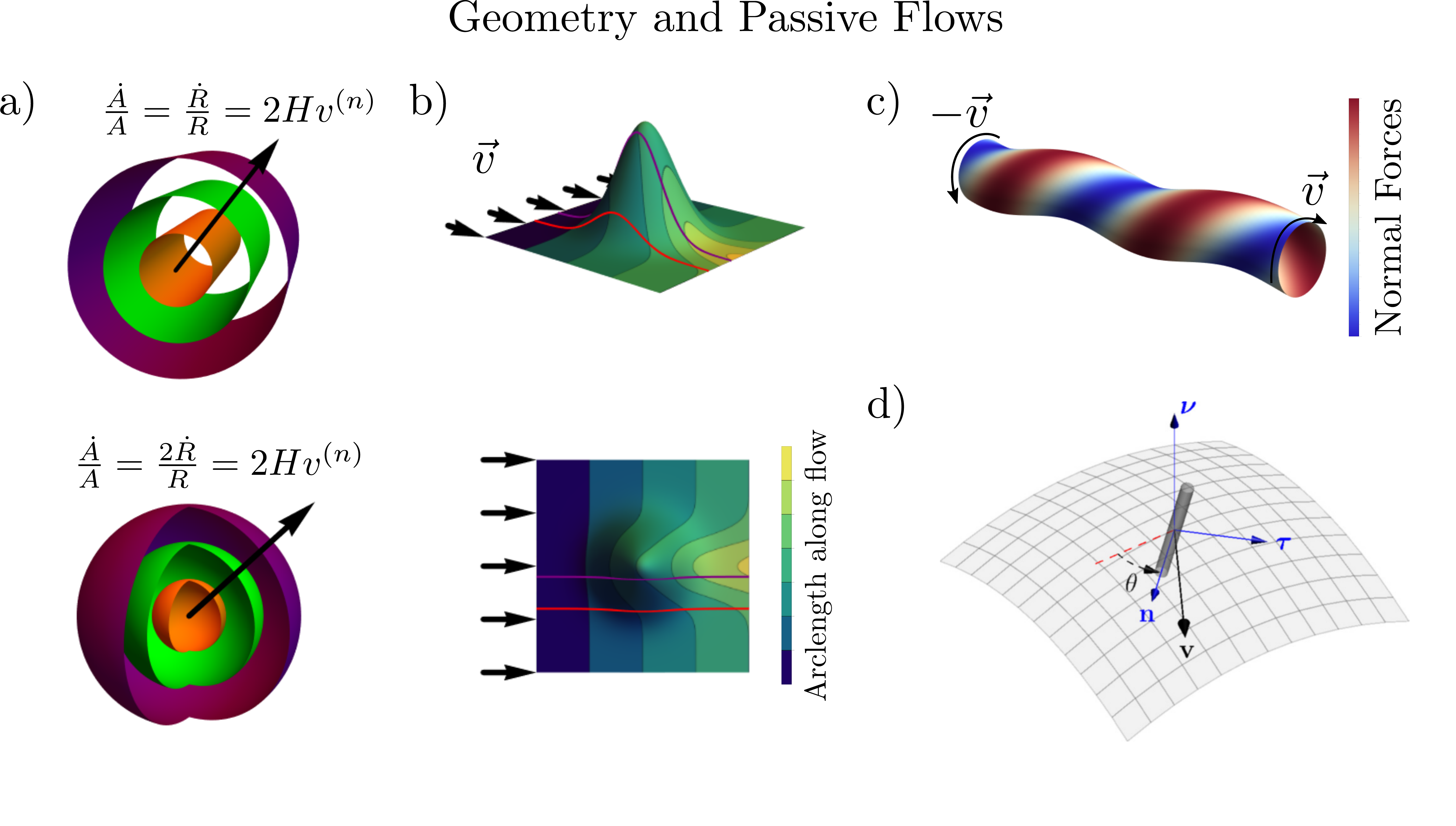}
\caption{\textbf{Interplay between geometry and passive flows.} a) Area changes arise due to normal velocities on curved surfaces (a sphere and a cylinder are shown). For incompressible fluids, this leads to a source density proportional to the mean curvature, $H$. b) Surfaces with Gaussian curvature require neighbouring fluid elements to travel paths of different lengths, which induces shear stresses independent of velocity gradients. c) Gradients in the angular velocity along a tube couple to curvature and lead to forces in the normal direction that can cause helical modes to grow. d) Schematic showing how the local frame induced by a nematic director sits on a curved surface, the texture couples to the curvature via projections of a full $3$D liquid crystal theory to a $2$D submanifold. Panel d) reproduced with permission from \cite{napoli_hydrodynamic_2016}.}
\label{fig:Fig2}
\end{figure*}

\subsection{Passive Surfaces}
Because the thickness of epithelial sheets is typically small compared to the lengthscales of deformations, such tissues are amenable to so-called ``thin film'' expansions, the type of which are common in continuum mechanics of elastic shells and fluid membranes \cite{landau_theory_1986,efrati_metric_2013}. Mathematically speaking, the idea is to approximate a tissue's geometry by a $2$D manifold (embedded in $\mathbb{R}^3$) and a scalar thickness, acting normal to the manifold.

For example, consider representing the epithelium by the thin film expansion of an elastic shell of thickness $t$, with constitutive relation for the stress given by $\sigma^{ij}=\mathcal{A}^{ijkl}\epsilon_{kl}$ (where $\epsilon_{ij}$ is the strain tensor and the components of $\mathcal{A}^{ijkl}$ are elastic moduli). The shape of the tissue at mechanical equilibrium is given by minimising the following free energy
\begin{align}\label{eq:shellEnergy}
	\mathcal{F}= &\frac{t}{8}\int_{\mathcal{M}}\left(g_{ij}-\bar{g}_{ij}\right)\mathcal{A}^{ijkl}\left(g_{kl}-\bar{g}_{kl}\right)\mathrm{d}A_{\mathcal{M}}\nonumber\\ 
	&+ \frac{t^3}{24} \int_{\mathcal{M}}\left(b_{ij}-\bar{b}_{ij}\right)\mathcal{A}^{ijkl}\left(b_{kl}-\bar{b}_{kl}\right)\mathrm{d}A_{\mathcal{M}},
\end{align}
where $g_{ij}$ and $b_{ij}$ are the metric and extrinsic curvature tensors, respectively, of the $2$D manifold, $\mathcal{M}$.  $\bar{g}_{ij}$ and $\bar{b}_{ij}$ are the so-called target metric and curvature tensors, respectively, and can be thought of as extensions of the notion of a ``rest length'' from elasticity theory. The form of $\mathcal{F}$ immediately demonstrates a non-trivial, but perhaps unsurprising result. Mechanical equilibrium depends not just on the intrinsic geometry of the tissue--- \textit{i.e.}, how it is stretched, compressed or sheared in the plane tangent to the surface--- but also on its extrinsic quantities, such as the curvatures that are encoded by $b_{ij}$.  In other words, although $\mathcal{M}$ is manifestly $2$D, the energy associated with its deformation accounts for finite thickness of the tissue via a thin-film expansion, and hence imbues it with a finite bending moment (or energetic cost of curvature).

The passive response of epithelial sheets is not exclusively elastic, however, and typically mimics that of a viscous fluid on intermediate timescales due to cell intercalation, division, and apoptosis.  In this context, geometry also has an important role to play.  Although, the interplay between flow and curvature is no longer a consequence of implicit thickness, as it is in the case of elasticity.  Rather, it is dependent on the notions of \textit{compressibility} and \textit{shear}, which are, in turn, related to the rate of deformation tensor, $d_{ij}$.  In non-flat geometries $d_{ij}$ is given by \cite{arroyo_relaxation_2009}
\begin{equation}
	d_{ij} = \frac{1}{2}\left(\nabla_iv_j +\nabla_jv_i\right) - b_{ij}v^{(n)},
	\label{eq:d}
\end{equation}
where $v_i$ are the tangential components of the velocity of the surface, $v^{(n)}$ is the normal component, and $\nabla_i$ is the $i$-th component of the covariant derivative.

To see how geometry and compressibility are linked, consider the extreme example of an  incompressible fluid, which must satisfy the constraint $\mathrm{Tr}(d)= {d_i}^i=0$.  In a flat geometry, this is just $\nabla_i v^i = 0$, which is the same as Eq.~(\ref{eq:incomp}) (but written in component form). In curved geometries, however, the condition of incompressibility is modified to
\begin{equation}
	\nabla_iv^i =2Hv^{(n)},
	\label{eq:incomp_geom}
\end{equation}
where $H = \mathrm{Tr}(b)/2$ is the mean curvature (\textit{i.e.}, the arithmetic mean of the two principal curvatures). To see why, think of a cylinder or sphere expanding outward along their radial direction at velocity $\dot R$ (Fig.~\ref{fig:Fig2}a). Because of their curvature, the non-zero normal velocity results in an increase in surface area, which implies that there must be a source term in the equations, since the tissue is incompressible. The required flux must equal the rate of change of area (per unit area)--- \textit{i.e.}, $\dot A/A$. It is relatively simple to convince oneself that this is equal to $2Hv^{(n)}$ in both cases.

To see how geometry is related to shear, it is helpful to first recall that, at low Reynolds number, inertia can be ignored. Therefore, force balance implies the Stokes equation--- \textit{i.e.}, the divergence of shear stresses are equal to gradients in pressure, or $\eta\,\nabla_i {d^i}_j = \nabla_j\,p$. On a \textit{curved} surface, this unpacks to
\begin{equation}
	\eta\left(\Delta + K \right)v_i = \nabla_i p,
	\label{eq:modified_Stokes} 
\end{equation}
where $\Delta=\nabla_i \nabla^i$ is the B\"{o}chner Laplacian, and $K = \mathrm{det}({b_i}^j)$ is the Gaussian curvature (\textit{i.e.}, the product of the two principal curvatures). Here, it is the second term on the left-hand-side that is of interest.  That is, a non-zero Gaussian curvature can give rise to forces due to viscous dissipation that are proportional to $\vec{v}$, and independent of shear gradients. Such terms have been shown to be important in the diffusion of conical proteins in lipid membranes \cite{morris_mobility_2015,morris_signatures_2017,daniels_curvature_2016} and in the hydrodynamics of liquid crystals on curved surfaces \cite{pearce_geometrical_2019}. A heuristic understanding of the origins of this term can be gained by considering a fluid flow over a `bump' (Fig.~\ref{fig:Fig2}b). Imposing a constant velocity along a given boundary, we see immediately that the fluid travelling over the centre of the bump must travel a greater distance than the fluid whose flow lines are off-centre.  This means that the bump cannot be negotiated unless there is relative movement of neighbouring fluid elements--- \textit{i.e.}, there must be fluid \emph{shear}. Eq.~(\ref{eq:modified_Stokes}) tells us that the forces arising from this additional curvature-induced shear are proportional to velocity and Gaussian curvature. As an aside, we note that the actual flow lines of an incompressible fluid must also take account of Eq.~(\ref{eq:incomp_geom}), since a bump also has mean curvature.

A further interplay between geometry and flows can occur in cylindrical geometries where gradients in azimuthal shear can lead to helical deformations of cylindrical shells (Fig.~\ref{fig:Fig2}c). In the case of elastic materials, such deformations can lead to the formation of plectonemes \cite{audoly_elasticity_2010} and helical instabilities \cite{al-izzi_shear-driven_2020}. Crucial to this phenomena is the fact that viscous forces from velocity gradients have components outside the tangent plane of the manifold (\textit{i.e.},~in the normal direction). This means that tangential flows can generate motion in the normal direction on curved surfaces \cite{rahimi_curved_2013,al-izzi_shear-driven_2020}. The dimensionless number associated with these out-of-plane forces is often called the Scriven-Love number \cite{scriven_dynamics_1960,sahu_geometry_2020}.

In addition to the aforementioned isotropic effects, it is also possible to extend the theories of ordered nematic fluids to curved and deformable surfaces. In this case, there arises a coupling between the nematic order parameter, $\vec{n}$ (where $\vec{n}=-\vec{n}$ and $|\vec{n}|=1$) and curvature, such that filaments will rotate in order to minimise the energy associated with their deformations in $\mathbb{R}^3$ \cite{napoli_extrinsic_2012,napoli_hydrodynamic_2016} (Fig.~\ref{fig:Fig2}d). Conversely, nematic order can also be used to \textit{generate} curvature, with the surface bending to reduce in-plane stresses generated by a given texture \cite{griniasty_curved_2019}. The ratio between nematic (or Frank) stresses and bending stress can be interpreted in a similar way to the F\"{o}ppel-von K\'{a}rm\'{a}n number in the classical elasticity of shells \cite{griniasty_curved_2019,audoly_elasticity_2010}.

We conclude this section by remarking that the interplay between curvature and flows makes finding stable numerical methods to solve hydrodynamic theories on curved manifolds a non-trivial challenge (at least beyond simplified axisymmetric cases). It is worth noting, however, that significant progress on this front has been made recently using mixed Lagrangian-Eulerian approaches \cite{torres-sanchez_modelling_2019,sahu_arbitrary_2020}, unfitted finite element methods \cite{barrett_stable_2016} and isogeometric analysis \cite{vasan_mechanical_2020}.

\subsection{Active Surfaces}

\begin{figure*}[!htbp]
\center\includegraphics[width=0.9\textwidth]{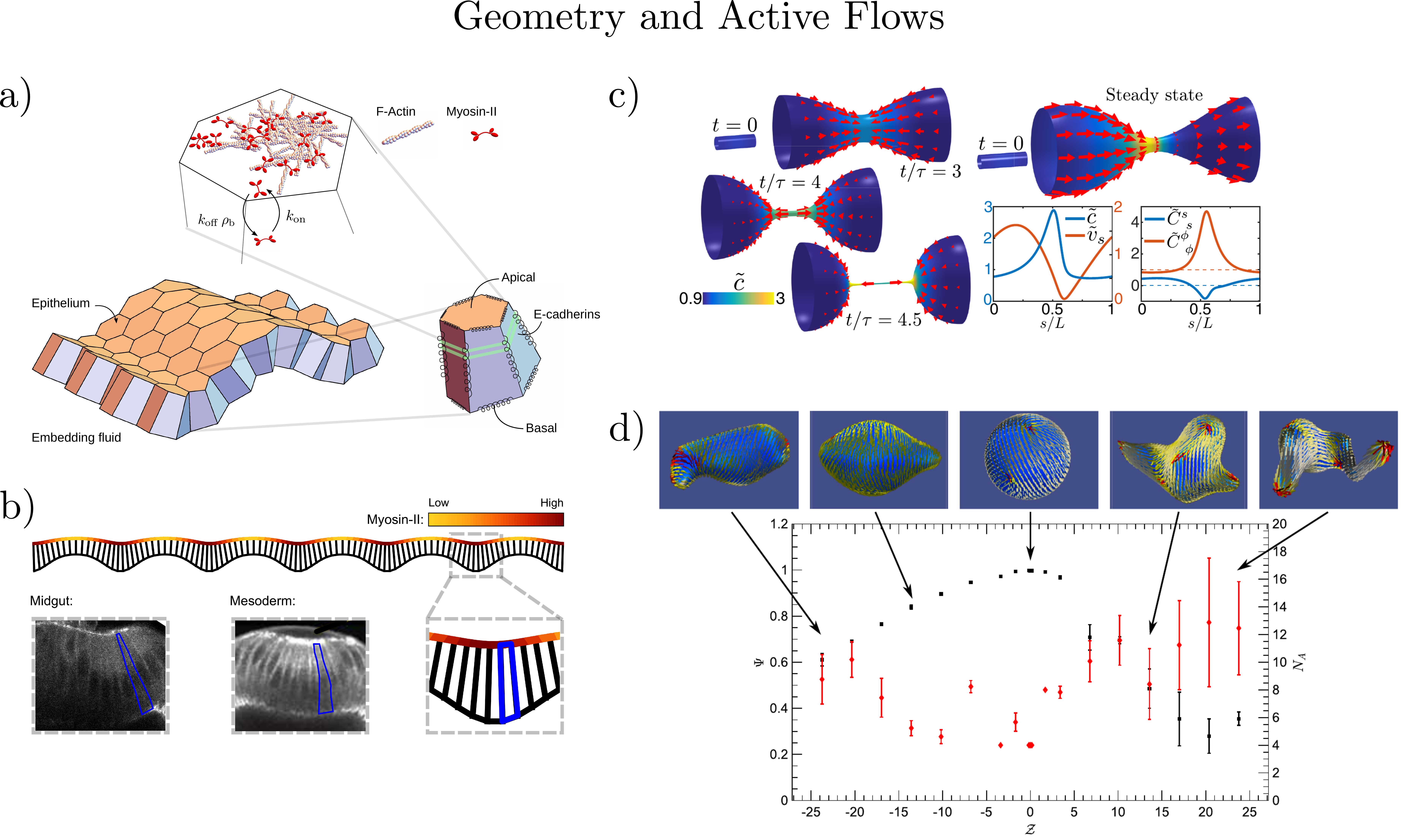}
\caption{\textbf{Interplay between geometry and active flows.} a) Schematic of myosin-II and actin at the apical surface of an epithelial cell. b) Example of a bending and thickening instability of a model of an epithelia driven by local increases in the concentration of myosin. The contraction of the apical side due to increased myosin leads to an active bending moment and concomitant thickening of the epithelium in order to preserve local volume. Mesoderm inset reproduced with permission from \cite{conte_biomechanical_2012}. c) Active hydrodynamic instabilities on a cylinder driven by some localised increase in chemical concentration, $\tilde{c}$. The right-hand side also shows the steady state shape along with plots of the velocity field ($\tilde{v}$) and concentration ($\tilde{c}$) along with the principle curvature components, $C^{s}_{s}$, $C^\phi_\phi$. Reproduced with permission from \cite{mietke_self-organized_2019}. d) The morphodynamics of an active liquid-crystalline shell surrounded by a viscous fluid for a range of values of dimensionless activity, $\mathcal{Z}$, where $\mathcal{Z}<0$ ($\mathcal{Z}>0$) is contractile (extensile). $\Psi$ is the sphericity of the vesicle. Reproduced with permission from \cite{metselaar_topology_2019}.}
\label{fig:Fig3}
\end{figure*}

The role of active forces in shaping embryos was first discussed in the modern context by Odell \textit{et.al.}~in 1981 where the idea of a simple excitatory chemical network driving active forces via the change in rest length of the apical faces of an epithelium was introduced \cite{odell_mechanical_1981}. Although this simple model could reproduce many shapes seen across a variety of developmental processes it was limited to  $2$D and did not model the underlying biochemical processes explicitly relying instead on a generic excitatory/inhibitory mechanism.

There are many ways in which a passive theory can be made active, these include growth driven activity \cite{goriely_mathematics_2017,goriely_five_2018} and coupling a passive elastic shell model via the rest metric and cruvature [\textit{cf}.~Eq.~(\ref{eq:shellEnergy})] to some underlying chemical reaction equation \cite{miller_geometry_2018}. The original work of Odell \cite{odell_mechanical_1981}, modern models of \textit{Drosophila} gastrulation \cite{allena_simulation_2010} and \textit{Volvox} inversion \cite{hohn_dynamics_2015,haas_elasticity_2015} fall into the later category (often not even modelling the active processes explicitly).

A second possibility is to follow a theorist's approach.  That is, write down all the terms allowed by symmetry arguments.  This leads to a very general theory with active tensions, bending moments and chiral torques \cite{salbreux_mechanics_2017,naganathan_active_2014,naganathan_actomyosin-driven_2016,pimpale_cell_2020}. 

There is however, still the challenge of relating such broad approaches to the underlying biology.  This is embodied by an apparently simple question: how can tangent-plane active stresses--- such as contractility, for example--- give rise to dynamics in the normal direction?

One way is via the mechanics of closed surfaces, like cylinders, for example \cite{mietke_self-organized_2019}. Here, local increases in chemical concentrations can lead to active stresses which change the equilibrium shape of the $2$D surface, \textit{i.e.}~shrinking the radius (Fig.~\ref{fig:Fig3}c). Such theories can also produce complex pulsatile dynamics and steady states \cite{mietke_self-organized_2019}.

Another way is to consider more specific models inspired by epithelia. For example, we have already seen how thin-film expansions link forces in the tangent plane to bending moments in the mechanical equilibrium of passive elastic shells [\textit{cf.}~Eq.~(\ref{eq:shellEnergy})]. It is relatively straightforward to extend these ideas to active descriptions of elastic-like epithelia, where contractile forces at apical faces give rise to bending moments. Moreover, a further coupling to curvature can be generated by explicitly imposing a local volume constraint \cite{morris_active_2019,haas_nonlinear_2019}. For example, by expanding the cell volume in the thickness $t$, giving  
\begin{equation}
	V = t\sqrt{\mathrm{det}(g_{ij})}\left(1-t H\right) + \mathcal{O}\left(t^3\right).
\end{equation}
If $V$ is a constant (\textit{i.e.}, if cell volume is conserved, for example) this allows for a coupling between the metric, the tissue thickness and the curvature. That is, a constriction of the apical faces can lead to both thickening of the baso-lateral faces, and enlargement of the basal surface (Fig.~\ref{fig:Fig3}a \& b) \cite{bailles_genetic_2019,morris_active_2019}.

Next, we turn our attention to ordered fluids with activity. On static curved surfaces the dynamics of topological defects is intimately coupled to curvature.  In particular, the Gaussian curvature of the surface dictates spontaneous defect production in extensile active nematics \cite{ellis_curvature-induced_2017,pearce_geometrical_2019,pearce_defect_2020}. The interplay between defects and curvature becomes even more important when the surface is allowed to deform. In such systems, both topological defects and activity control the morphology of protrusions \cite{metselaar_topology_2019,keber_topology_2014}, as shown in Fig.~\ref{fig:Fig3}d. It has recently been proposed that these defects can be used to engineer active metamaterial surfaces \cite{pearce_programming_2020,warner_curvature_2010,modes_mechanical_2012,modes_blueprinting_2011}. Although such theories have been explored extensively, their relevance for real biological systems has yet to be investigated in detail. Promisingly, it was shown recently that topological defects in tissue cultures dictated the location of extruding cells \cite{saw_topological_2017}.  It is therefore possible that defects may play and important role in generating out of plane deformations and morphologies in development. 

As with active hydrodynamics in the flat setting, it is currently challenging to solve the full non-linear equations numerically and fit these theories to experimental data. In the final section we will discuss possible approaches and frameworks which may provide some way forward on this front.

\section{Discussion}

Throughout this short review we have highlighted some of the challenges associated with comparing current (covariant) theories of active hydrodynamics with data. This not only includes inferring model parameters, but also their dependence on underlying biological processes. Here we will discuss the present outlook and how we might better use experimental data and quantitative analysis methods to inform theory.

Firstly, the challenge of fitting parameters to partial differential equations is, in general, ill-posed, with many possible solutions. Recently, however, approaches based on Bayesian inference have made some progress in fluid dynamics and shape-shifting elastic bilayers \cite{cotter_bayesian_2009,hoang_determining_2014,van_rees_growth_2017}. It is currently an open challenge to see if these methods can be extended to biologically realistic scenarios, such as embryogenesis.

Nevertheless, within the past $10$--$15$ years, there has been significant advances in the quantitative methods used to analyse tissue dynamics \cite{merkel_triangles_2017,serra_dynamic_2020,guirao_unified_2015,graner_discrete_2008,marmottant_discrete_2008}. Generally speaking, this appears likely to continue, in no small part due to the improvements in machine learning driven segmentation algorithms and other computational tools which will help make developmental biology more high throughput \cite{aigouy_epyseg_2020,stringer_cellpose_2021,etournay_tissueminer_2016}.

Of particular note are methods concerned with how deformations of discrete objects, namely cells, contribute to coarse-grained tissue-level properties. For example, shear rates at the tissue level can arise from local topological rearrangements of the cells. These can occur in three main ways: cell division, where a single cell splits into two daughter cells; cell death (apoptosis), and; so-called T$1$ transitions \cite{merkel_triangles_2017,blanchard_tissue_2009,tlili_colloquium_2015,guirao_unified_2015,popovic_active_2017,wang_anisotropy_2020}. So far, such methods have been used in a variety of situations to gain insight into the material deformation rates of tissues, both \textit{in-vitro} \cite{tlili_migrating_2020} and \textit{in-vivo} \cite{merkel_triangles_2017,guirao_unified_2015,blanchard_tissue_2009}.

Indeed, these techniques have also yielded some progress in estimating the values of passive hydrodynamic coefficients \cite{tlili_migrating_2020,popovic_inferring_2020} without recourse to the aforementioned Bayesian inference methods. However, it remains a significant challenge to disambiguate between the passive rheological properties of the tissue and active forces, at least without making significant unsubstantiated assumptions.  This goes back to a point made in Sec.~\ref{sec:hyd}: active coefficients can be arbitrary functions of the other variables that describe the system, and are therefore hard to determine without extensive biological controls. The problem therefore manifests as one of experimental characterisation, and whether genetic and/or other manipulations can be designed to make the desired inference(s) possible.

Geometry presents further problems. For example, it is an open question how far such data analysis methods can be extended to the curved surfaces and complex tessellation patterns in real embryos (\textit{i.e.},~Scutoids \cite{gomez-galvez_scutoids_2018}, for example).  Not to mention the non-trivial challenge of unambiguously reconstructing $3$D data from $2$D slices \cite{sharp_inferring_2019}.

Here, we suggest that the relatively new field of Discrete Differential Geometry (DDG) may provide inspiration~\cite{crane_glimpse_2017,berlin_discrete_2008}. DDG is a mathematical framework that has had a profound impact on computer graphics and digital geometry processing. The aim is to take definitions and concepts from classical differential geometry and apply them to discrete geometrical objects. This often leads to a ``no free lunch'' scenario where, for example, various equivalent definitions of the continuous curvature of a contour in $2$D are now different in the discrete setting (\textit{e.g.}~turning angle vs.~length variation vs.~osculating circle) and only preserve a subset of the properties of of the continuous definitions \cite{crane_glimpse_2017}. DDG is then focussed on developing mathematics and algorithms to translate all the tools from classical differential geometry to this discrete setting and understand the features which are preserved, and thus, what the best choice of definition is for a given task.

We argue that making use of such methods in tissue mechanics may be prove fruitful in interpolating between discrete models of cells such as vertex models \cite{rauzi_physical_2013,fletcher_vertex_2014,bi_motility-driven_2016,tetley_tissue_2019,rozman_collective_2020}, continuum models and real experimental systems. In fact, recently some progress was made in coarse graining a simple discrete model with differential tensions to a full continuum theory using ideas very much in the spirit of DDG \cite{haas_nonlinear_2019}.

In conclusion, we hope that this topical review has provided some insight into the notion of flows on curved and deformable manifolds, and their applicability in developmental biology and morphogenesis.  There remain many challenges, ranging from developing analytical tools and numerical methods, to better integration of such approaches with real biological data.  We therefore welcome work in the area.

\section*{Declarations}
The authors have no competing interests to declare.

\acknowledgements{RGM and SCA-I acknowledge funding from the EMBL-Australia program. The authors thank J.~Binysh (University of Bath, UK) for critical reading of the manuscript.}

\end{document}